\documentstyle[aps]{revtex}
\def\gtrsim
{\raise2pt\hbox
{$\displaystyle{}\mathrel{\mathop>_{\raise1pt\hbox{$\sim$}}}{}$}} 
\def\lesssim
{\raise2pt\hbox
{$\displaystyle{}\mathrel{\mathop<_{\raise1pt\hbox{$\sim$}}}{}$}}
\begin{document}
\title{
Electronic pressure on ferromagnetic domain wall
} 
\author{Gen { Tatara}\footnote{corresponding author: e-mail  
tatara@ess.sci.osaka-u.ac.jp, address: Graduate School of Science, 
Osaka University, Toyonaka, Osaka 
560-0043, Japan,
Fax +81-6-6850-5494, Phone 
+81-6-6850-5544}  \\
Graduate School of Science, Osaka University, Toyonaka, Osaka 
560-0043, Japan\\
and Yasuhiro { Tokura}$^{1}$\\
$^{1}$
NTT Basic Research Laboratories, 3-1 Wakamiya Morinosato, Atsugi, Kanagawa
243-0198, Japan
}
\maketitle
\begin{abstract}
The scattering of the eletron by a domain wall in a nano-wire is 
studied perturbatively to the lowest order. 
The correction to the thermodaynamic potential of the electron system due 
to the scattering is
calculated from the phase shift. 
The wall profile is determined by taking account of this correction, 
and the result indicates that the 
wall in a ferromagnet with small exchange coupling can be 
squeezed to be very thin to lower the electron energy.  
\end{abstract}
%%%%%%%%%%%%%%%
\sloppy
%%%%%%%%%%%%%%%%%%%

%\section{Introcduction}
Recently the magnetoresistance in various magnetic 
nano-structures such as multilayers, wires and contacts has been 
attracting growing interest. 
The magnetoresistance effects in such systems are induced by the 
spatial variation of the localized magnetic moments. 
Hence, to grasp the physics of the resistance due to a domain 
wall scattering is fundamental to an understanding of the mechanism of
the magnetoresistance.
The theoretical study of this problem goes back to more than 25 years 
ago by Cabrera and Falicov\cite{Cabrera74}. 
The scattering mechanism by the wall is due to the non-adiabatic deviation 
of the electron spin from the local magnetization. In 
the case of  a thick wall with thickness $\lambda \gg k_{F}^{-1}$ 
($k_{F}$ being the Fermi wave length), the electron spin can 
adiabatically follow the spatial rotation of the magnetization as it 
traverses the wall, resulting in a 
negligibly small reflection. 
Cabrera and Falicov thus concluded that the resistivity contribution 
due to domain walls is important only in a limit of 
$k_{F}\lambda\rightarrow0$.
This limit was unrealistic at that time, where 3$d$ metals in 
the bulk were considered mostly.  
However, recent technology has put the problem in the limelight, 
firstly by the direct measurements of domain wall 
resistance\cite{Hong96,Otani97,Ruediger98} and secondly 
by fabrication of a very small domain wall in nano-scale 
contacts\cite{Garcia99,TZMG99}.
Motivated by the recent experiments detailed theoretical studies 
including the quantum correction in the dirty 
case\cite{TF97,Geller98}, spin-dependent lifetime\cite{Zhang97}, 
etc. have been 
made\cite{TZMG99,Yamanaka96,vanHoof99,Imamura00,Nakanishi99}.
These investigations revealed that the problem was not as simple as 
had been believed. In fact theories predict that even the sign of 
the domain wall contribution to the resistivity depends on temperature, 
system size, amount of disorder as well as the material, and 
experimental results indeed show
 such diversity.
A clarifying explanation of the whole behavior observed in wires
has not given so far.
In nano-scale contacts, the situation seems to be simpler.
There the wall is believed to be constrained in a small contact 
region of an atomic scale due to a rapid change of the contact 
shape, and thus 
adiabaticity does not 
hold, resulting in strong scatterings\cite{vanHoof99,TZMG99,Bruno99}.
A recent experiment on Ni contacts revealed a magnetoresistance of 
about 300\% at room temperature\cite{Garcia99}, which is 
significantly large compared with other giant magnetoresistive 
materials such as multilayers. 
The result has been interpreted as due to a strong electron scattering 
by a domain wall, which is trapped in the narrow contact region in the 
absence of the magnetic field, and which is expelled by applying a 
magnetic field\cite{TZMG99}. 

The strong scattering by the wall in nano magnets suggests that the 
electronic state is modified. This possibility is studied by 
looking at the phase shift, $\phi\equiv {\rm Arg}\;t$, $t$ being the 
transmission coefficient due to the domain wall scattering.
This phase shift is related to the total energy of the electron 
system through the equality 
$D=\frac{1}{\pi}\frac{d\phi}{d\epsilon}$\cite{DMB69,Avishai85},
where $D$ is the density of states and $\epsilon$ is the energy.
Thus the stability of the electron system in the presence of the 
domain wall can be discussed in term of the phase shift.

The 
relation between the scattering and the stability of the electronic 
state has been studied in the non-magnetic case. 
Experimentally it was shown in 
a Au contact that the 
tensile force of the contact as the contact is 
stretched is related with the 
conductance and becomes maximum when the conductance shows a plateau 
with a quantized value of integer times $2e^2/h$\cite{Rubio96}.
This result indicates that the electron system is stabilized when the 
conductance takes a quantized value.
Such behavior has been explained considering the scattering by the 
potential energy in the constriction during the prolongation 
and taking account of the energy 
shift of the electron system, which is calculated from the scattering 
phase shift by use of the above relation\cite{Stafford97}.

In the case of ferromagnetic wires and contacts, 
a domain wall may affect the 
stability of the electrons and thus the wall structure can be modified 
to lower the electron energy.  
The essential parameter of the wall which 
governs the transmission 
coefficient is the thickness, $\lambda$. In this paper 
the thermodynamic potential of the electrons is calculated as a 
function of $\lambda$. This thickness is determined to minimize the 
total energy of the wall, i.e., the 
magnetostatic energy plus the electron energy. 
An interesting point is that in 3$d$ metals the energy scale of the electron, 
$\epsilon_{F}$, is much larger than that of the magnetization, the 
exchange ($J$) and anisotropy ($K$) energies. 
Thus even a small effect of the scattering of the electron might results in an 
energy shift comparable to the magnetostatic energy.
In fact it will turn out that this is the case in a 
ferromagnet with 
small $J$, and the wall would be 
compressed to be very thin, 
$\lambda\sim k_{F}^{-1}$. 

Considering a narrow wire, the system is treated as in one dimension. 
The calculation is at zero temperature.
The electron channels considered here 
are those which contribute to the 
transport and the quantized behavior of the conductance is not 
discussed here.
We consider the exchange 
interaction between the electron spin and the local spin 
and the Hamiltonian is

\begin{equation}
\label{1}H=\sum\limits_{k\sigma }\epsilon _k c_{k\sigma }^{\dagger
}c_{k\sigma }-g\int dx{\bf S}(x)(c^{\dagger }\mbox{\boldmath$\sigma$}c),
\end{equation}

where $\epsilon _k\equiv \hbar ^2k^2/2m-\epsilon _F$ ($\epsilon _F$ being
the Fermi energy). The spin index is denoted by $\sigma =\pm $ and 
$\mbox{\boldmath$\sigma$}$ is the Pauli matrix
(spin indices are suppressed in the second term).
Here, the local spin ${\bf S}$ has a spatial dependence of a domain wall.
In terms of a polar coordinate $\left( S_z\equiv S\cos \theta \right) $ the
wall is represented as $\cos \theta\rightarrow 1$ for $x\ll -\lambda$ 
and $\cos \theta\rightarrow-1$ for $x\gg\lambda$,
$\lambda $ being the thickness of the wall. 
To proceed we 
carry out a local gauge transformation in the electron spin space so that
the $z$-axis is chosen to be along the local direction of spin 
$\overrightarrow{S}$\cite{TF94}. The transformation is written as
$a_\sigma =\sigma \cos \left(\frac \theta 2 \right) c_\sigma 
-i\sin \left(\frac \theta 2 \right) c_{-\sigma}$, 
where the electron in the new frame is denoted by $a_\sigma $. The
Hamiltonian is modified to be
$
H=H_{0}+H_{\rm int}
$
where 
$H_{0}\equiv \sum\limits_{k\sigma }\epsilon _{k\sigma }a_{k\sigma }^{\dagger
}a_{k\sigma }$ and 
$\epsilon _{k\sigma }\equiv \epsilon _k-\sigma \Delta
$is the energy of uniformly polarized electron with the exchange splitting 
$\Delta \equiv g\left| {\bf S}\right|$. The term $H_{\rm int}$ describes the
interaction between the electron and the wall and is given by 
\cite{TF97}

\begin{equation}
\label{DWint}H_{\rm int}\equiv \frac{\hbar ^2}{2m}\frac 1L\sum\limits_{kq}\left[
-\left( k+\frac q2\right) A_qa_{k+q}^{\dagger }\sigma _xa_k+\frac 1{4L}%
\sum\limits_pA_pA_{-p+q}a_{k+q}^{\dagger }a_k\right] ,
\end{equation}
where

\begin{equation}
\label{6}A_q\equiv \int dxe^{-iqx}\nabla \theta,
\end{equation}
is a domain-wall form factor and $L$ is the system size.
Note that $\nabla \theta\rightarrow 0$ for $|x|\rightarrow \infty$ 
and so the interaction is localized near the wall. 
The 
spin-conserving process in the rotated frame as the electron goes 
through the wall corresponds to the adiabatic process, where the 
electron spin follows the rotation of the magnetization, and the spin 
flip corresponds to the deviation from adiabaticity.

The interaction $H_{\rm int}$ leads to a scattering of the electron 
and thus contributes to a resistivity due to the 
wall as has been discussed before based on a linear response 
theory\cite{TF97,TZMG99}.  
Here we study the phase shift due to the scattering,
calculating the $T$-matrix perturbatively with respect to $H_{\rm int}$.
The wave function corresponding to the incoming state with the energy 
$E$ and the spin $\sigma$, which we denote as $\Psi_{E}^{\sigma'\sigma}$ 
($\sigma'$ is the final state spin), is 
calculated as
\begin{equation}
	\Psi_{E}^{\sigma'\sigma}(x)=\delta_{\sigma',\sigma}e^{ik_{\sigma}x}
	+\int dx' G^0_{E\sigma'}(x-x') 
	\sum_{k'}e^{ik'x'}T_{k'\sigma',k_\sigma \sigma},
	\label{Psi1}
\end{equation}
where $G^0_{E\sigma'}(x-x') \equiv -i\frac{m}{|k_{\sigma'}|} 
e^{ik_{\sigma'}|x-x'|}$ is the free Green function and
$T_{k'\sigma',k_\sigma \sigma}\equiv 
<k'\sigma'|T|k_{\sigma}\sigma>$, 
$T\equiv H_{\rm int} \frac{1}{1-G^0_{E}H_{\rm int}}$.

The transmission and reflection coefficients are defined from the 
asymptotic behavior as
\begin{eqnarray}
	\Psi_{E}^{\sigma'\sigma}(x) &\rightarrow & \left\{
	\begin{array}{cc}
   \delta_{\sigma'\sigma}e^{ik_{\sigma}x}+r_{\sigma'\sigma}e^{-ik_{\sigma'}x} 
   &  (x\rightarrow -\infty) \\
     t_{\sigma'\sigma}e^{ik_{\sigma'}x} 
   &  (x\rightarrow +\infty)  \end{array} \right.   .
   	\label{rtdef}
\end{eqnarray}
The result is
\begin{eqnarray}
	t_{\sigma'\sigma}(k_{\sigma}) & = & 
	\delta_{\sigma'\sigma}-i\frac{mL}{|k_{\sigma'}|} 
	T_{k_{\sigma'}\sigma',k_{\sigma}\sigma} \nonumber\\
	r _{\sigma'\sigma}(k_{\sigma}) & = & -i\frac{mL}{|k_{\sigma'}|} 
	T_{-k_{\sigma'}\sigma',k_{\sigma}\sigma} .
	\label{tr}
\end{eqnarray}
By use of these 
coefficients, the resistance due to the wall is 
calculated based on the Landauer's formula (with two-terminals) as 
$R_w=R-R_0$, where 
\begin{equation}
R^{-1}=(e^2/h)\sum_{\sigma}(|t_{\sigma\sigma}|^2 
+\frac{k_{F,-\sigma}}{k_{F\sigma}}|t_{-\sigma,\sigma}|^2), \label{Rw}
\end{equation}
and 
$R_0\equiv (h/2e^2)$ is the resistance without the wall.

The diagonal (spin $\sigma$) component of the 
transmission coefficient for the incoming wave vector of $k_{\sigma}$
is calculated as
\begin{equation}
	t_{\sigma\sigma}(k_{\sigma})=1+i\frac{1}{8k_{\sigma}L}\sum_{q} |A_{q}|^2 
	\frac{3k_{\sigma}^2+k_{-\sigma}^2+2k_{\sigma} q} 
	{(q+k_{\sigma})^2-k_{-\sigma}^2-i0}.
	\label{t}
\end{equation}
The diagonal component of the phase shift, which is defined as 
$\phi_{\sigma}(\epsilon)\equiv 
{\rm Im}\ln t_{\sigma\sigma}(k)$ 
($\epsilon\equiv (\hbar^2 k^2/2m)$), is obtained as
\begin{equation}
	\phi_{\sigma}(\epsilon) =\frac{1}{8kL}{\rm P}
	\sum_{q}|A_q|^2 
	\frac{4k^2+2kq-2\sigma\tilde{\Delta}}{2kq+q^2+2\sigma\tilde{\Delta}},
	\label{phi}
\end{equation}
where $\tilde{\Delta}\equiv (k_{F+}^2-k_{F-}^2)/2$ and
P denotes taking the principal value. 
In the adiabatic limit of large $k_F\lambda$, 
the phase shifts of two spin channels acquire
Berry's geometrical phases\cite{Berry84}. The phase shift
difference of the two channel become $\pi$ since the
solid angle of the rotation of local magnetization
is $\pi$.

\begin{figure}
%\begin{center}
%\epsfxsize=8cm              
%\epsfbox{DW.eps}
%\end{center}
%
%
\caption{The lowest order contribution of the wall to the electron 
energy. The solid and wavy lines denote the electron and the wall, 
respectively.
\label{FIGdiag}}
\caption{The function $f(x)$. Solid line is the result of a 
perturbative calculation ((eq. (\ref{fdef})) and dashed lines are the 
non-perturbative (numerical) results for $\zeta=0.3$ and $\zeta=0.8$.
It is seen that $f$ does not 
depend much on $\zeta$, and the perturbative approximation is fairly good.
The region $k_{F}\lambda<1$, which is not physically meaningful, 
is plotted by dotted lines.
\label{FIGf}}
\caption{
The total energy due to a domain wall, $E_{\rm w}/\epsilon_{F}$,  
as a function of thickness, $\lambda$. Solid and dashed lines are 
for $\zeta=0.3$ and $0.8$, respectively, and three lines correspond to 
$J/\epsilon_{F}=0.02$, $0.01$ and  $0.001$ from the top to the bottom. 
It is seen that the energy of the electron system is lowered for a very 
thin wall, $k_{F}\lambda\sim1$, if $J/\epsilon_{F}\lesssim 0.001$ 
($J/\epsilon_{F}\lesssim 0.02$) 
for $\zeta=0.3$ ($\zeta=0.8$), respectively.   
\label{FIGEw}}
\end{figure}
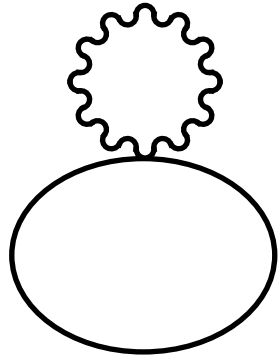
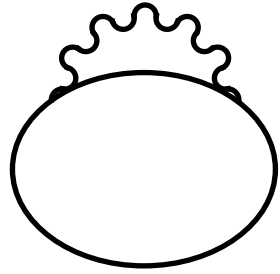
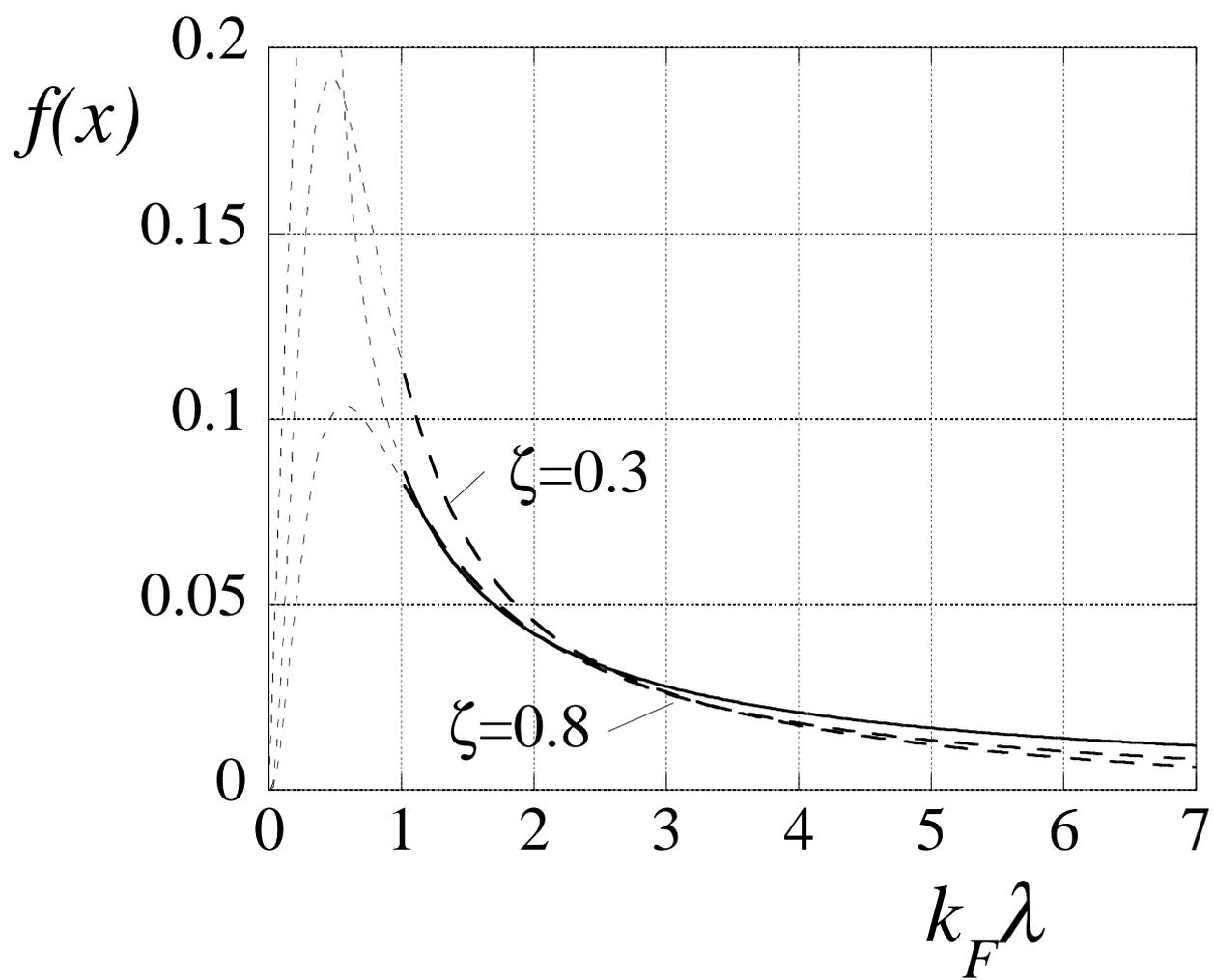
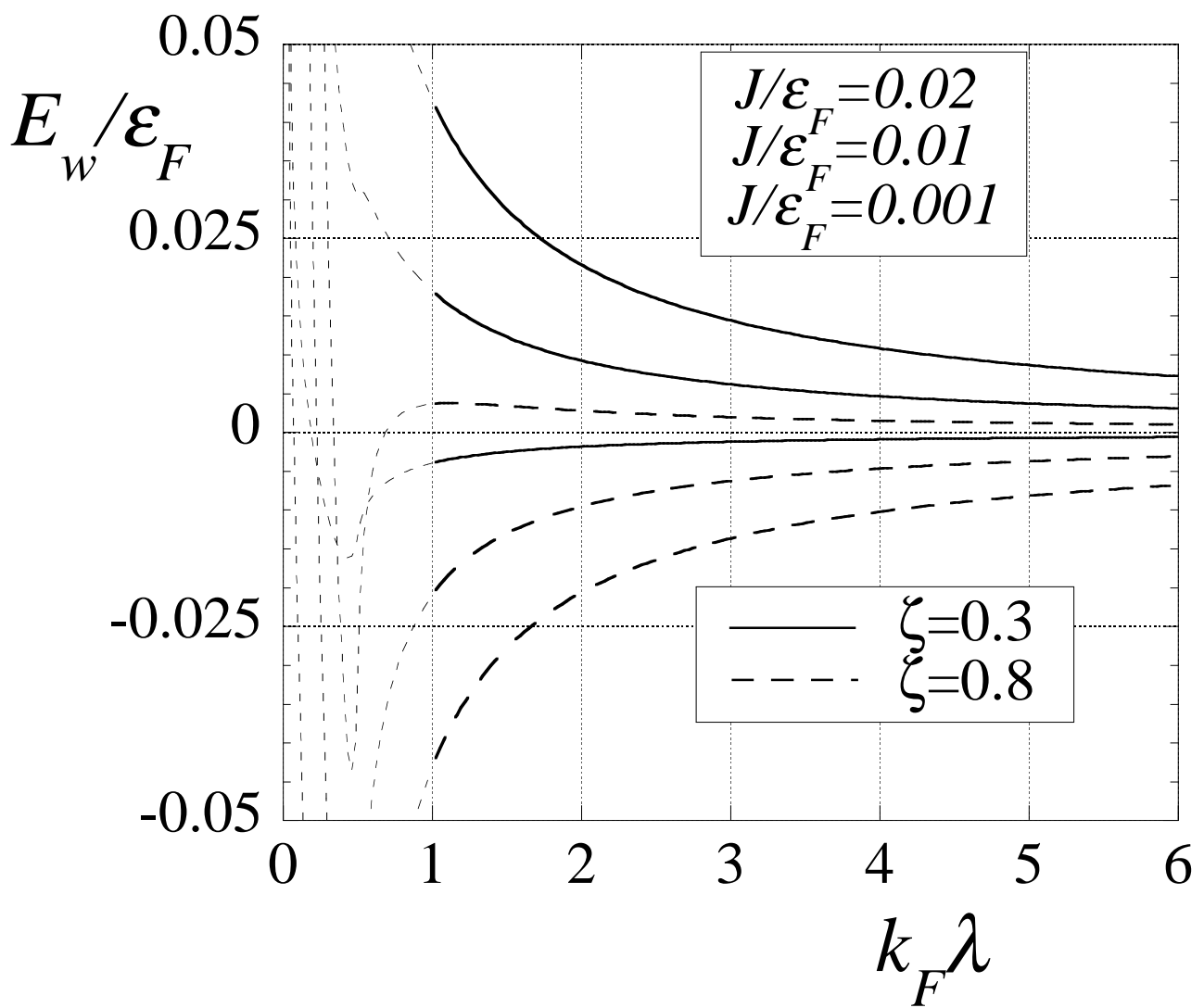

The phase shift contributes to the change of the density of state of 
the electron according to the relation
$\delta D_{\sigma}(\epsilon) 
=(1/\pi)(d\phi_{\sigma}(\epsilon)/d\epsilon)$\cite{DMB69,Avishai85}.
Thus the thermodynamic potential of the electron system 
(defined as $\Omega_{\rm e}\equiv 
-\frac{1}{\beta}\ln{\rm Tr}e^{-\beta(H-\epsilon_{F}N)}$) 
is shifted due to the scattering
by an amount of
\begin{eqnarray}
	\delta \Omega_{\rm e}&=& \sum_{\sigma}\int_0^{\epsilon_{F\sigma}} 
	d\epsilon (\epsilon-\epsilon_{F\sigma})\delta D_{\sigma}(\epsilon)
	\nonumber\\
	&=& -\frac{1}{\pi} \sum_{\sigma}\int_0^{\epsilon_{F\sigma}} 
	d\epsilon \phi_{\sigma}(\epsilon)+\delta\Omega_{0},
	\label{dEdef}
\end{eqnarray}
where $\Omega_{0}\equiv \frac{1}{\pi} 
\sum_{\sigma}\epsilon_{F\sigma}\phi_{\sigma}(\epsilon=0)$ is an 
irrelevant constant (independent of domain wall parameter), which we 
neglect. 
Thus we obtain 
\begin{equation}
	\delta \Omega_{\rm e}= -\frac{1}{8m}\frac{1}{L^2}\sum_{q} |A_{q}|^2 
	\sum_{k\sigma} f_{k\sigma}
	\frac{4k^2+2kq-2\sigma\tilde{\Delta}}{2kq+q^2+2\sigma\tilde{\Delta}},
	\label{delOm0}
\end{equation}
where $f_{k\sigma}=\theta(\epsilon_{F}-\epsilon_{k\sigma})$ is the Fermi 
distribution function at zero temperature ($\theta(x)$ is the step 
function).
We note that this expression of $\delta \Omega_{\rm e}$ is also obtained
directly by evaluating the diagrams described in 
Fig. \ref{FIGdiag}\cite{Mahan90}.
In fact these contributions are calculated as
\begin{equation}
	\delta \Omega_{\rm e}= -\frac{1}{8m}\frac{1}{L^2}\sum_{q} |A_{q}|^2 
	\sum_{k\sigma} f_{k\sigma}\left[\frac{\frac{2}{m}(k+\frac{q}{2})^2} 
	{\epsilon_{k+q,-\sigma}-\epsilon_{k\sigma}} -1 \right],
	\label{delOm}
\end{equation}
which is equal to (\ref{delOm0}).

The expression (\ref{delOm0}) reduces after $k$-summation to
\begin{equation}
	\delta \Omega_{\rm e}= -\frac{1}{8\pi m L} \sum_{q} 
	\frac{(k_{F+}^2-k_{F-}^2)}{q^2} |A_{q}|^2 
	\left[-(k_{F+}-k_{F-}) 
	+\frac{(k_{F+}^2-k_{F-}^2)}{2q}
	\ln\left|
	\frac{q+k_{F+}+k_{F-}}{q-(k_{F+}+k_{F-})}
	\right|\right].
	\label{delOm2}
\end{equation}
To proceed, we approximate the profile of the wall to simplify the 
calculation. 
In the case of an ideal wall determined solely by the exchange and 
uniaxial 
anisotropy, the profile is $\cos\theta=\tanh(x/\lambda)$ and 
so $A_{q}=\pi/\cosh(\pi q \lambda/2)$.
By noting that the essential 
feature of the wall is that the 
form factor $A_{q}$ is large only for $|q|\leq \lambda^{-1}$, we here  
replace 
$A_{q}=2$ for $|q|\leq \lambda^{-1}$ and $A_{q}=0$ for 
$|q|\geq \lambda^{-1}$.
(We have checked that this approximation does not change the result very much.)
By this approximation the energy shift is calculated as
\begin{equation}
	\delta \Omega_{\rm e}= -\epsilon_{F}\zeta^2 f(k_{F}\lambda),
	\label{delOm3}
\end{equation}
where 
\begin{equation}
	f(x)\equiv\left(\frac{4}{\pi^2}\right)
	\left( x+\frac{1}{4}(1-4x^2) \ln\left|\frac{2x+1}{2x-1}\right| 
	\right),
	\label{fdef}
\end{equation}
and 
\begin{equation}
	\zeta=\frac{k_{F+}-k_{F-}}{k_{F+}+k_{F-}}.
\end{equation}
The minus sign of $\delta\Omega_{\rm e}$ indicates that the electron 
system is stabilized in the case of a thiner wall ($\lambda k_{F} 
\simeq 1$) which causes a strong scattering. 
This is because the two spin channels are mixed by the domain wall. 
The function $f(x)$ grows at smaller $x$ as plotted in Fig. \ref{FIGf}.
The result (\ref{fdef}) of the perturbative calculation is not 
justified at $k_{F}\lambda \lesssim 1$ .
We thus estimated the function $f(x)$ non-perturbatively
by calculating the transmission amplitude by
the recursive Green's function method\cite{Ando91}.
We have used discretized form of the Hamiltonian 
Eq.(\ref{1}) with a tight-binding approximation.
The result of function $f$ is potted in Fig. \ref{FIGf} as dashed lines. 
The function $f$ turns out to depend on $\zeta$ only very little 
in the region physically meaningful 
(i.e., $k_{F}\lambda\gtrsim1$), which is 
consistent with the perturbative result (\ref{fdef}).
It is seen that the 
perturbative result (solid line) gives a good approximation to the 
full result. 

In the case of an insulating magnet the structure of the domain wall 
is determined solely by 
minimizing the magnetostatic energy.
The magnetostatic energy is written as  
\begin{equation}
	E_{\rm M}=\int \frac{dx}{a}  \left( 
	\frac{Ja^2}{2}(\nabla\theta)^2+\frac{K}{2}\sin^2\theta \right),
	\label{EMdef}
\end{equation}
where $J$ and $K$ are the exchange and anisotropy energies per site, 
respectively and $a$ is the lattice spacing  
(we assume that the anisotropy energy $K$ 
includes the contribution from the dipole energy).
In the ideal case the wall thickness is given by 
$\lambda_{0}\equiv \sqrt{J/K}a$. 
Here we consider the case where the wall 
is thinner compared with the bulk case, $\lambda \ll \lambda_0$, where 
the effect of the electron can be important.
We assume that the cross-section of the wire is only weakly 
varying in the length scale we consider, and 
thus neglect the effect from the wire geometry discussed in ref. 
\cite{Bruno99}. 
The magnetostatic energy of the wall is thus dominated by the exchange 
term of eq. (\ref{EMdef}). 
By approximating the wall as $\nabla\theta=\pi/(2\lambda)$ 
for $|x|\leq \lambda$ and $\nabla\theta=0$ otherwise, the magnetostatic energy
is obtained as
\begin{equation}
	E_{\rm M}\simeq \int dx 
	\frac{Ja}{2}(\nabla\theta)^2 \simeq \frac{\pi^2}{4}\frac{a}{\lambda}J.
	\label{EM1}
\end{equation}
In the metallic case the total energy due to the wall is given by the 
sum of the electron 
part and the magnetic part ($E_{\rm w}=\delta \Omega_{\rm 
e}+E_{\rm M}$):
\begin{equation}
	E_{\rm w}=\frac{\pi^2}{4}\frac{a}{\lambda} J
	-\epsilon_{F}\zeta^2 f(k_{F}\lambda).
	\label{ew}
\end{equation}
It is seen that the electron energy and magnetostatic energy compete 
with each other. 
The behavior of $E_{\rm w}$ is plotted in Fig. \ref{FIGEw}.
Only the region of very small $\lambda$ is plotted since for larger 
$\lambda$ 
the functions vary monotonically.

The parameter which determines the relative strength of the 
two energies is
\begin{equation}
	\eta\equiv 0.1\times \frac{4\epsilon_{F}\zeta^2}{\pi^2 J },
	\label{etadef}
\end{equation}
where a factor of 0.1 is to take into account the value of the 
function $f(x)$ at $x\sim1$.
The value of $\eta$ exceeding 1 indicates that the wall energy can be 
lowered if the wall becomes  very thin. 
In 3$d$ transition metals such as Ni, $\epsilon_{F}\sim 4$eV, $J\sim 0.1$eV 
and $\zeta\sim 0.3$, and thus $\eta\sim 0.15$ in the case of $d$-band. 
Therefore the electron part 
is not strong enough to modify the domain wall structure. As is seen 
from the figure, in the case of $\zeta=0.3$, if $J$ is as small as 
$\sim 0.001\epsilon_{F}$, the wall is compressed to become an atomic 
size, $k_{F}\lambda\sim1$.
In a strong ferromagnet with $\zeta=0.8$ the crossover value below which 
the electron energy wins occurs 
around $J\sim 0.16\epsilon_{F}$.

If $J$ is smaller than the threshold value, the wall is very thin and 
so leads to an almost perfect 
reflection $|t|\sim 0$ for channels below the threshold and 
$|t|\sim1$ for those above. In other words, the wall is self-organized 
to give rise to a 
quantized value of the conductance, just as the wire shape 
does during the prolongation in non-magnetic wires\cite{Stafford97}. 
In recent experiment of Ni contacts clear quantization of the conductance
in unit of $2e^2/h$ has been observed under zero and small magnetic 
field\cite{Ono99}. A possible existence of a domain wall has been 
suggested there, but naively domain wall in nano-contact is expected to 
lead to a deviation from the conductance quantization since it gives rise 
generally to a transmission probability of $T\neq 1$. 
The present mechanism might be relevant to the clear quantization 
observed,
since suppression of the effective exchange coupling $J_{\rm 
eff}$ is expected in nano-contacts due to low-dimensionality. 
(The suppression of $J_{\rm eff}$ is indicated in 
experiments on films as a decrease of ferromagnetic transition 
temperature ($T_{c}$) as the thickness is reduced\cite{Huang94}.) 

In ref. \cite{Bruno99}, the importance of the wire shape on the wall 
structure has been demonstrated.
This effect is due to the shape anisotropy and so the relevant energy 
scale is $J$ or $K$. Hence the compression of the wall by the 
electron system would be more important in ferromagnetic wire 
with small $J$ or large $\zeta$. 
We have calculated in one-dimension but 
the present result would qualitatively hold in the bulk case as well. 

Suitable system for 
experiments would be a low $T_c$ ferro- or 
ferri-magnetic materials. 
The present study indicates that in the metallic case
there can be a crossover value of 
$J$, below which $\lambda$ becomes of order of lattice spacing.
From the point of view of 
controlling the electron density, experiments on 
magnetic semi-conductors might be interesting.

%\section{Summary}
To summarize,
the energy shift of the electron system by the scattering by a domain 
wall has been calculated. The result indicates that the wall lowers the 
energy of the electron system and the effect is larger for a thiner wall 
than a thick one. 
This contribution thus competes with the 
magnetic exchange energy, which favors a thick wall.  
In 3$d$ metals in the bulk the effect turns out not to be strong  
enough to affect the wall structure, but it can dominate 
in the case with smaller $J$ or a larger splitting or in nano-wires, 
in which cases the wall is squeezed by 
the electron system to be about an atomic scale. 

\section*{Acknowledgements}
G.T. thanks H. Matsukawa for useful discussion. 
He also thanks Max Planck Institut f\"{u}r Mikrostrukturphysik for its 
hospitality during his stay at an early stage of this work and 
Alexander von Humboldt Stiftung for financial support.
This work is supported by a Grand-in-Aid for Scientific 
Research from the Ministry of Education, Science, 
Sports and Culture.
Y.T. acknowledges financial support from the Dutch Organization for
Fundamental Research on Mater (FOM) and from the NEDO joint research
program (NTDP-98).
 
%%%%%%%%%%%%%%%%%%%%%%%%%%%%%%%%

%
%%%%%%%%%%%%%%%%%%%%%%%%%%%%%%%%
%
\end{document}